\documentclass{osa-article}

\journal{oe}

\articletype{Research Article}

\begin{document}
\title{Instant ghost imaging: algorithm and on-chip implementation}

\author{Zhe Yang, Wei-Xing Zhang, Yi-Pu Liu, Dong Ruan, and Jun-Lin Li\authormark{*}}

\address{State Key Laboratory of Low-dimensional Quantum Physics and Department of Physics, Tsinghua University, Beijing 100084, China}

\email{\authormark{*}center@mail.tsinghua.edu.cn} 



\begin{abstract}
Ghost imaging (GI) is an imaging technique that uses the correlation between two light beams to reconstruct the image of an object. Conventional GI algorithms require large memory space to store the measured data and perform complicated offline calculations, limiting practical applications of GI. Here we develop an instant ghost imaging (IGI) technique with a differential algorithm and an implemented high-speed on-chip IGI hardware system. This algorithm uses the signal between consecutive temporal measurements to reduce the memory requirements without degradation of image quality compared with conventional GI algorithms. The on-chip IGI system can immediately reconstruct the image once the measurement finishes; there is no need to rely on post-processing or offline reconstruction. This system can be developed into a realtime imaging system. These features make IGI a faster, cheaper, and more compact alternative to a conventional GI system and make it viable for practical applications of GI.
\end{abstract}

\section{Introduction}

Ghost imaging (GI) is an imaging technology which reconstructs the image of an object by calculating the correlation between two beams (test and reference).
The test beam interacts with the object and is collected by a bucket detector without spatial resolution, and the reference light field is detected using a space-resolving detector without going through the object.
It has been demonstrated that correlations of both quantum-entangled \cite{Shih1} and thermal light sources \cite{Boyd1, Lugiato1, Lugiato2, Shih2} can be used to achieve the GI.
The image can be formed without a lens (lensless ghost imaging) \cite{LGI_Cao, LGI_Shih, LGI_Lorenzo} or only by using a single-pixel detector (computational ghost imaging, CGI) \cite{2008CGI, 3D2013, 3D2016}.
Due to the underlying physics and potential applications in many fields, including lidar \cite{2016Lidar}, tomography \cite{2018Tomography}, and medical imaging \cite{2016Xray, 2016Xray2, 2018Xray}, GI has attracted much attention in recent years \cite{2013BackGI, 2015few, 2019BGI, Zhe1, Optica1, DeepGI}.
It has also been extended to different domains with certain freedoms of correlation, including atomic domain \cite{2016atom, 2019atom}, time domain \cite{2016time, 2016Ctime, 2019Colortime}, and spiral imaging \cite{2014OAM, 2017OAM}.

A significant obstacle to practical applications of GI is that reconstituting an image requires massive temporal measurements, which necessitates huge memory space with high space complexity. This limitation stems from conventional GI algorithms. For example, the background subtraction algorithm requires the second-order correlation function calculation \cite{Lugiato1, Lugiato2, 3D2013, 2018Xray}
\begin{equation}
{G}(x) = \left\langle {(S - \left\langle S \right\rangle )[I(x) - \left\langle {I(x)} \right\rangle ]} \right\rangle,
\label{GI}
\end{equation}
where $ \left\langle {\; \cdot \;} \right\rangle  = {{(1} \mathord{\left/
 {\vphantom {{(1} N}} \right.
 \kern-\nulldelimiterspace} N})\sum\nolimits_{n = 1}^N {( \cdot )}$ means the ensamble average over $N-$times measured signals, ${{I}(x)}$ is the intensity in certain position $x$ of the reference beam, and $S$ is the bucket signal of the test beam. Calculation of this algorithm is time-consuming and post-processed offline, antithetical to online or realtime computation. There have been many attempts to improve the imaging quality of GI, such as differential ghost imaging (DGI) \cite{2010DGI, DGItime}, iterative ghost imaging \cite{IterGI01, IterGI02}, and higher-order ghost imaging \cite{Higher01, Higher02}. However, few works attempted to reduce the memory required or the space complexity to implement online GI.
 
Compressive sensing \cite{Compressive01, Compressive02}, a convex optimization procedure, reduces the required number of acquisitions for GI with good image quality \cite{Compressive03, Compressive04, 10hz}. However, it is at the cost of more computing resources, which increases the GI's dependence on the computer. The single-pixel imaging uses a complete orthogonal basis, such as the Fourier basis \cite{2015NC, 11hz} and the Hadamard basis \cite{Hadamard, 1000hz}, to obtain a perfect image of the object. This solution requires complete sampling, for example, a $256\times 256$ image requires 65,536 measurements. Therefore, a computer is needed to do the inverse transform to obtain the image especially for the large images. To date, on-chip GI has not been perfected due to high space complexity.

We first proposed a sequence differential-based GI algorithm in 2015. It uses the signals between two consecutive temporal measurements, the $(n+1)^{th}$ and $n^{th}$, in the test and the reference beams, ${S_{n + 1}} - {S_n}$ and ${I_{n + 1}}(x) - {I_n}(x)$, to reconstruct the image of the object \cite{2016Bo,2018Zhuan,2019Zhuan}. Jun-Lin Li introduced the algorithm into computer-based experiments as sequence differential ghost imaging (SDGI) in his doctoral dissertation in 2016 \cite{2016Bo}. Ya-Xin Li et al. also discussed the virtually identical algorithm with computer-based experiments in detail \cite{2019Copy}. However, the strong dependence on the computer is an obstacle to practical applications of GI, and previous related works \cite{2016Bo,2018Zhuan,2019Zhuan,2019Copy} have not solved this problem. In this work, to demonstrate the validity and the hardware feasibility of the SDGI algorithm, we developed a prototype on-chip hardware system using a single field-programmable gate array (FPGA), without any external memory; it can process 500 measurements per second online. This system was named instant ghost imaging (IGI) due to one significant advantage of this system: its image reconstruction time is almost zero and the image is formed immediately once the temporal measurement is complete. The on-chip IGI system makes GI computer-independent for the first time.

IGI offers the following advantages:
\begin{itemize}
\item IGI can drastically reduce memory requirements and space complexity without increasing computation.
\item IGI does not reduce image quality compared to the background subtraction algorithm.
\item IGI is a generalized GI algorithm that can be used for lensless ghost imaging and CGI.
\item The on-chip IGI hardware system measures the signal and reconstructs the image online: it does not rely on post-processing or offline reconstruction.
\item The structure of the on-chip IGI system is compact and much smaller than the computers needed to calculate the correlation function in conventional GI procedures. Moreover, the IGI hardware system could be developed into a realtime imaging system at a frame rate of more than 24 frames per second. These features make IGI a faster, cheaper, and more compact alternative to a conventional GI system and make it viable for practical applications of GI.
\end{itemize}

\section{Methods}
\subsection{Instant ghost imaging algorithm}

Experimentally, we can use the $N$-times measurements to calculate Eq. (\ref{GI}) of the background subtraction algorithm
\begin{equation}
{G}(x) = \frac{1}{N}\sum\limits_{n = 1}^N {{S_{n}} \cdot {I_{n}}(x) - \frac{1}{N}\sum\limits_{n = 1}^N {{S_n}}  \cdot } \frac{1}{N}\sum\limits_{n = 1}^N {{I_n}(x)},
\label{eq:GI_N}
\end{equation}
where the bucket signal $S$ of the test beam is given by ${S} = \int {{I}({x_t})} T({x_t})d{x_t}$, ${{I}({x_t})}$ is the intensity of the test beam, and $T({x_t})$ is the transmissivity function of the object.

The IGI algorithm we proposed differs from Eq. (\ref{eq:GI_N}) in using $(N+1)$ measurements
\begin{equation}
G^{IGI}(x) = \frac{1}{{2N}}\sum\limits_{n = 1}^N {({S_{n + 1}} - {S_n})[{I_{n + 1}}(x) - {I_n}(x)]},
\label{eq:IGI}
\end{equation}
where ${S_{n + 1}}\;-\;{S_n}$ and ${I_{n + 1}}(x) - {I_n}(x)$ are the temporal differential signals between two consecutive measurements of the bucket detector and the reference detector.

We can demonstrate that Eq. (\ref{eq:IGI}) of the IGI algorithm is equivalent to Eq. (\ref{eq:GI_N}) of the background subtraction algorithm when $N$ is rather large.
It can be inferred that Eq. (\ref{eq:IGI}) has four terms
\begin{equation}
\begin{array}{l}
{G^{IGI}}(x) = \frac{1}{{2N}}\sum\limits_{n = 1}^N {{S_{n + 1}} \cdot {I_{n + 1}}(x)}  + \frac{1}{{2N}}\sum\limits_{n = 1}^N {{S_n} \cdot {I_n}(x)} \\
\;\;\;\;\;\;\;\;\;\;\;\;\;\;\; - \frac{1}{{2N}}\sum\limits_{n = 1}^N {{S_{n + 1}} \cdot {I_n}(x) - \frac{1}{{2N}}\sum\limits_{n = 1}^N {{S_n} \cdot {I_{n + 1}}(x)} }.
\end{array}
\label{eq:IGI_N}
\end{equation}

When $N$ is rather large, it can be assumed that
\begin{equation}
\begin{array}{*{20}{l}}
{\left\langle {S \cdot I(x)} \right\rangle  = \frac{1}{N}\sum\limits_{n = 1}^N {{S_n} \cdot {I_n}(x)}  \approx \frac{1}{N}\sum\limits_{n = 1}^N {{S_{n + 1}} \cdot {I_{n + 1}}(x)} }\\
{\left\langle S \right\rangle  = \frac{1}{N}\sum\limits_{n = 1}^N {{S_n}}  \approx \frac{1}{N}\sum\limits_{n = 1}^N {{S_{n + 1}}} }\\
{\left\langle {I(x)} \right\rangle  = \frac{1}{N}\sum\limits_{n = 1}^N {{I_n}(x) \approx } \frac{1}{N}\sum\limits_{n = 1}^N {{I_{n + 1}}(x)} }.
\end{array}
\label{eq:Average}
\end{equation}

According to the theory of quantum optics, the temporal coherence and spatial coherence of thermal light are very short, and the fluctuation of thermal light is statistically irrelevant \cite{book}. Therefore, two successive thermal light measurements are independent of each other.
Using the statistical law that $\left. {\left\langle {A\cdot B} \right.} \right\rangle  = \left\langle A \right\rangle\; \cdot\left\langle B \right\rangle$ when $A$ and $B$ are independent random variables,
the last two terms of the Eq. (\ref{eq:IGI_N}) take the form of
\begin{equation}
\begin{array}{l}
\frac{1}{N}\sum\limits_{n = 1}^N {{S_{n + 1}}} \cdot{I_n}(x) = \frac{1}{N}\sum\limits_{n = 1}^N {{S_{n + 1}}}  \cdot \frac{1}{N}\sum\limits_{n = 1}^N {{I_n}(x)} \\
\frac{1}{N}\sum\limits_{n = 1}^N {{S_n}} \cdot{I_{n + 1}}(x) = \frac{1}{N}\sum\limits_{n = 1}^N {{S_n}}  \cdot \frac{1}{N}\sum\limits_{n = 1}^N {{I_{n + 1}}(x)}.
\end{array}
\label{eq:Average2}
\end{equation}

According to Eq. (\ref{eq:Average}) and (\ref{eq:Average2}), we find that Eq. (\ref{eq:IGI_N}), i.e. Eq. (\ref{eq:IGI}), is equal to Eq. (\ref{eq:GI_N}) when $N$ is rather large
\begin{equation}
G(x) \approx G^{IGI}(x).
\end{equation}
This requirement for $N$ is easy to satisfy because the number of measurements in GI is usually of the order of tens of thousands.

\subsection{Experimental setup}

\begin{figure*}[htb]
\centerline{\includegraphics[scale=0.35]{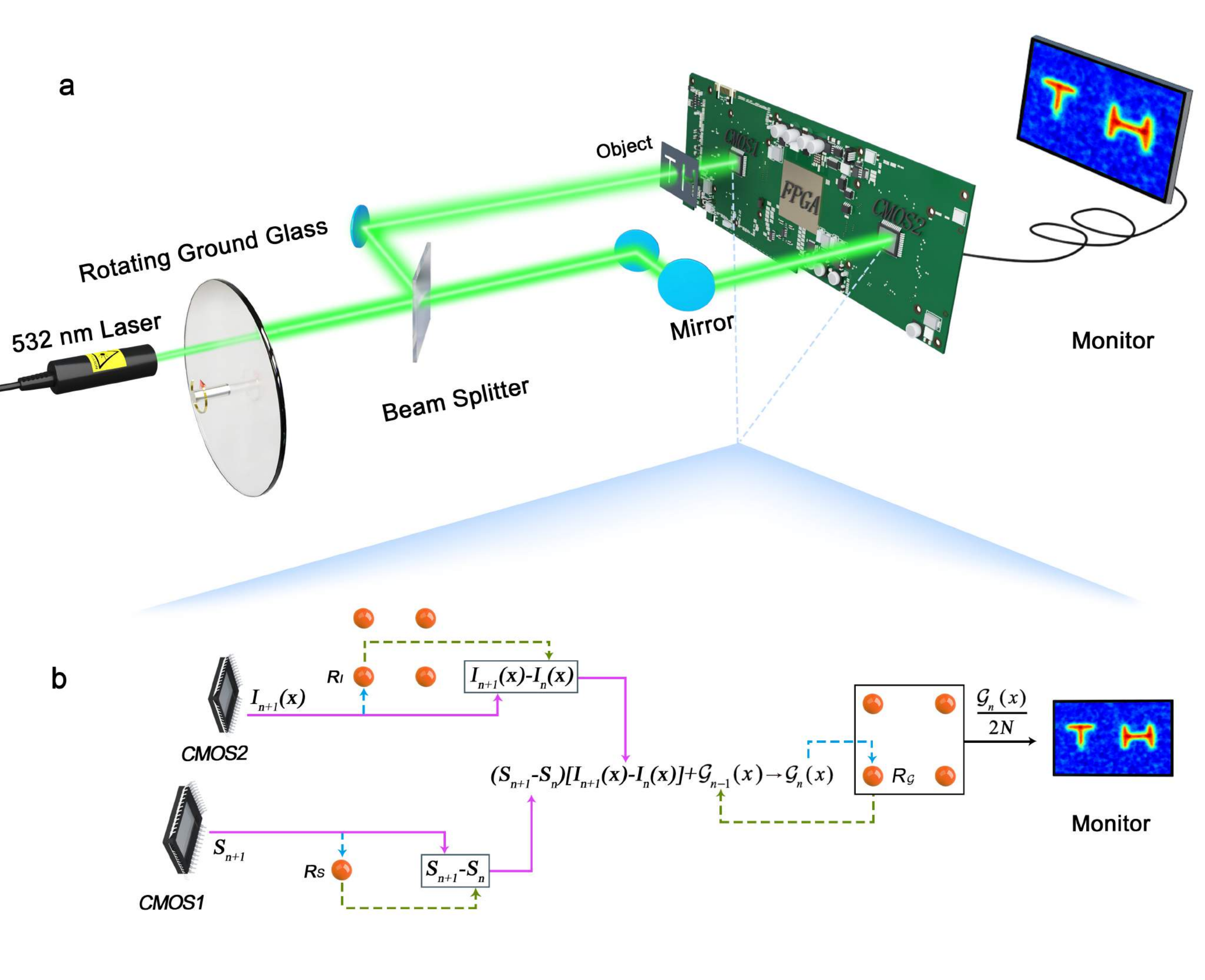}}
\caption{(a), Schematic of the experimental setup.
CMOS1, CMOS2: complementary metal-oxide semiconductor; FPGA: field-programmable gate array.
The pseudo-thermal light is produced by passing a 532 nm laser through a rotating ground glass disk and the light beam is split into two: one beam illustrates the object and is collected by CMOS1; the other is directly recorded by CMOS2. The FPGA-based on-chip IGI is used to reconstruct the image of the object using the IGI algorithm, and the intermediate results are shown in the monitor. The distances from the rotating ground glass disk to the object and to CMOS2 are both equal to 300 mm.
(b), The workflow of the IGI system. Green dashed lines: the FPGA extracts data from the corresponding register; Blue dashed lines: the FPGA stores new data in the register, overwriting old data. An orange ball represents a register unit; $S_n$ and  $I_n(x)$ are stored in the registers $R_S$ and $R_I$, and the intermediate result $\mathcal{G}_{n}(x) = ({S_{n + 1}} - {S_n})[{I_{n + 1}}(x) - {I_n}(x)]$ is stored in the register $R_{\mathcal{G}}$.}
\label{fig:figure_setup}
\end{figure*}

The schematic of the experimental setup is shown in Fig. \ref{fig:figure_setup}(a). A 532 nm laser light goes through a slowly-rotating ground glass disk to produce pseudo-thermal light. A beam splitter (BS) divides the light into two beams, the test beam and the reference beam. A binary mask object of letters TH is placed in the test beam 300 mm downstream of the disk. The mask is close to a complementary metal-oxide-semiconductor CMOS1  (PYTHON300) which is used to simulate the bucket detector; the bucket signal $S$ is calculated by summing up all the light intensities detected by the CMOS1. Another detector CMOS2 is in the reference beam at a distance of 300 mm from the ground glass disk. Each CMOS can carry out 500 measurements per second.
The hardware specification about the experimental setup can be found in the Methods section.

The entire calculation required for image reconstruction is performed in the IGI hardware, which consists of two CMOSs, an FPGA (Xilinx Kintex-7 XC7K325T) and a monitor. The FPGA is used to compute the temporal differential signals ${S_{n + 1}}-{S_n}$, ${I_{n + 1}}(x)-{I_n}(x)$, and their product $({S_{n + 1}} - {S_n})[{I_{n + 1}}(x) - {I_n}(x)]$; it can process all the 500 measurements per second made by each CMOS. 
The monitor shows the intermediate results of IGI for a fixed interval, typically four times per second. The IGI hardware system is completely on-chip because the two CMOSs, the FPGA, and the monitor are integrated on a printed circuit board (PCB). This results in a smaller and much more compact configuration than conventional GI systems. We also want to emphasize that the system contains only a single FPGA without any external memory.   

We now introduce the framework and workflow of the IGI hardware system, as shown in Fig. \ref{fig:figure_setup}(b). After the $ n^{th} $ measurement has been processed,  $S_n$, ${I_{n}}(x)$, and $\mathcal{G}_{n-1}(x)$, which is defined as $ {\cal G}_{n - 1}(x) = \sum\nolimits_{i = 1}^{n - 1} {({S_{i + 1}} - {S_{i}})[{I_{i + 1}}(x) - {I_{i}}(x)]}  $, are stored in the corresponding registers, $R_S$, $R_I$, and $R_{\mathcal{G}}$. When the $(n+1)^{th} $ signal is detected by two CMOSs, giving ${S_{n+1}}$ and ${I_{n+1}}(x)$, the FPGA can compute the differential signals ${S_{n + 1}} - {S_n}$ and ${I_{n + 1}}(x) - {I_n}(x)$, using ${S_{n}}$ and ${I_{n}}(x)$ from $R_S$ and $R_I$. $S_{n+1}$ and ${I_{n+1}}(x)$ overwrite $S_n$ and ${I_{n}}(x)$ in $R_S$ and $R_I$. $({S_{n + 1}} - {S_n})[{I_{n + 1}}(x) - {I_n}(x)]$ is then calculated and added to  $\mathcal{G}_{n-1}(x)$ to give $\mathcal{G}_{n}(x)$, which overwrites  $\mathcal{G}_{n-1}(x)$ in $R_{\mathcal{G}}$. This illustration of IGI workflow in processing one measurement shows that the on-chip IGI system can make a pair of measurements and process them immediately before the next measurement is made.
At every 125$^{th}$ measurements (i.e. four times per second), the monitor will show the intermediate result  $\mathcal{G}_{n}(x)/(2N)$. When the number of measurement $n$ increases to the preset $N$, the reconstructed image of the object is immediately available without any post-processing (hence the Instant in IGI).

\section{Results}
\subsection{Hanbury Brown and Twiss effect}

\begin{figure*}[htb]
\centerline{\includegraphics[scale=0.45]{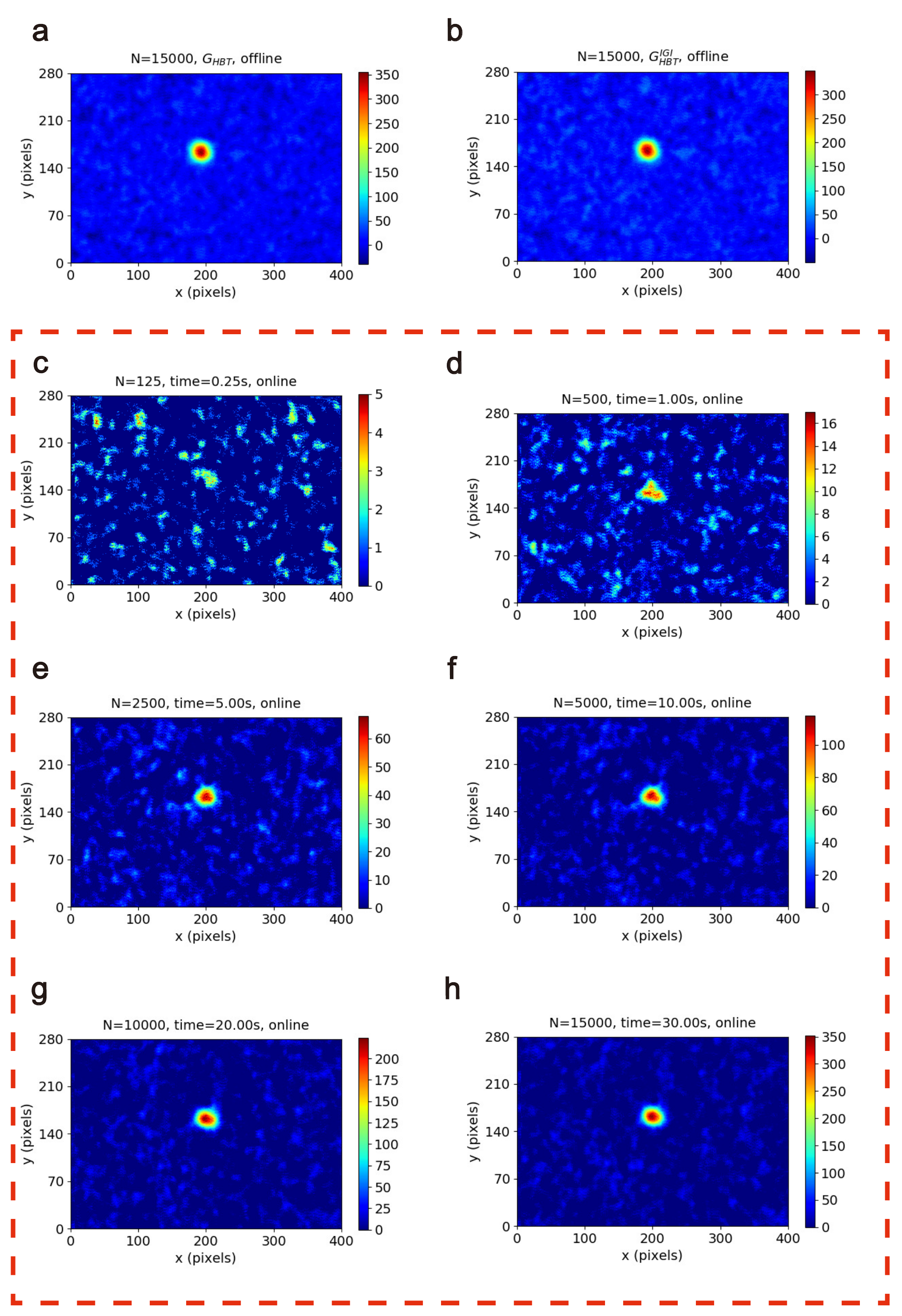}}
\caption{The Hanbury Brown and Twiss effect. The offline HBT effect from 15,000 measurements obtained (a) by the conventional $G_{HBT}$ algorithm; (b) by the $G_{HBT}^{IGI}$ algorithm. (c)-(g) The intermediate results and (h) the final result of the online IGI hardware system using the $G_{HBT}^{IGI}$ algorithm.} 
\label{figure_HBT}
\end{figure*}

GI is based on the second-order point-to-point correlation between the test beam and the reference beam. We demonstrate this correlation by conducting the Hanbury Brown and Twiss (HBT) experiment, which takes the form       
\begin{equation}
{G}_{HBT}({x_t},{x_r}) = \left\langle {[{I}({x_t}) - \left\langle {{I}({x_t})} \right\rangle ][{I}({x_r}) - \left\langle {{I}({x_r})} \right\rangle ]} \right\rangle,
\end{equation}
where  ${{I}({x_t})}$ and ${{I}({x_r})}$ are the light intensities detected by CMOS1 and CMOS2. 
We propose a new algorithm, based on the IGI algorithm      
\begin{equation}
\begin{array}{l}
G_{HBT}^{IGI}({x_t},{x_r}) = \frac{1}{{2N}}\sum\limits_{n = 1}^N {[{I_{n + 1}}({x_t}) - {I_n}({x_t})][{I_{n + 1}}({x_r}) - {I_n}({x_r})]}.
\end{array}
\end{equation}
It uses the differential signals of two beams to obtain the HBT effect. We have shown that this equation is theoretically equivalent to the HBT algorithm when $N$ is rather large. A detailed proof is given in the Appendix.

The HBT experiment is conducted on the setup shown in Fig. \ref{fig:figure_setup}(a) to verify the accuracy of the $G_{HBT}^{IGI}$ algorithm. The mask object is removed, and one pixel of the test beam is fixed, $x_t = x_{t0}$. The experiment is conducted using both offline and online methods. For the offline experiment, we take 15,000 measurements made at a rate of 25 measurements per second, store them in a computer, and use both the $G_{HBT}$ algorithm and the $G_{HBT}^{IGI}$ algorithm to process these data offline. The results, with image resolution 400$\times $280, are shown in Figs. \ref{figure_HBT}(a)-\ref{figure_HBT}(b). These two algorithms produce almost equal results.

For the online experiment, we use the on-chip IGI system to process the measured data at a rate of 500 measurements per second online. 
The results, showing the time and number of measurements, are shown in Figs. \ref{figure_HBT}(c)-\ref{figure_HBT}(h), which show that as time increases, the HBT effect becomes clearer. 
Note that when the time is 30 s, the 15,000 measurements have all been made and the final result is immediately available. 
The movie shown in the monitor of the IGI hardware system can be found in the \textcolor{urlblue}{Visualization 1}.

The experimental results show that the $G_{HBT}^{IGI}$ algorithm accurately calculates the second-order correlation for the two beams, thus providing a solid foundation for IGI to successfully reconstruct the image of the object.

\subsection{Image of the object}

\begin{figure}[htb]
\centerline{\includegraphics[scale=0.30]{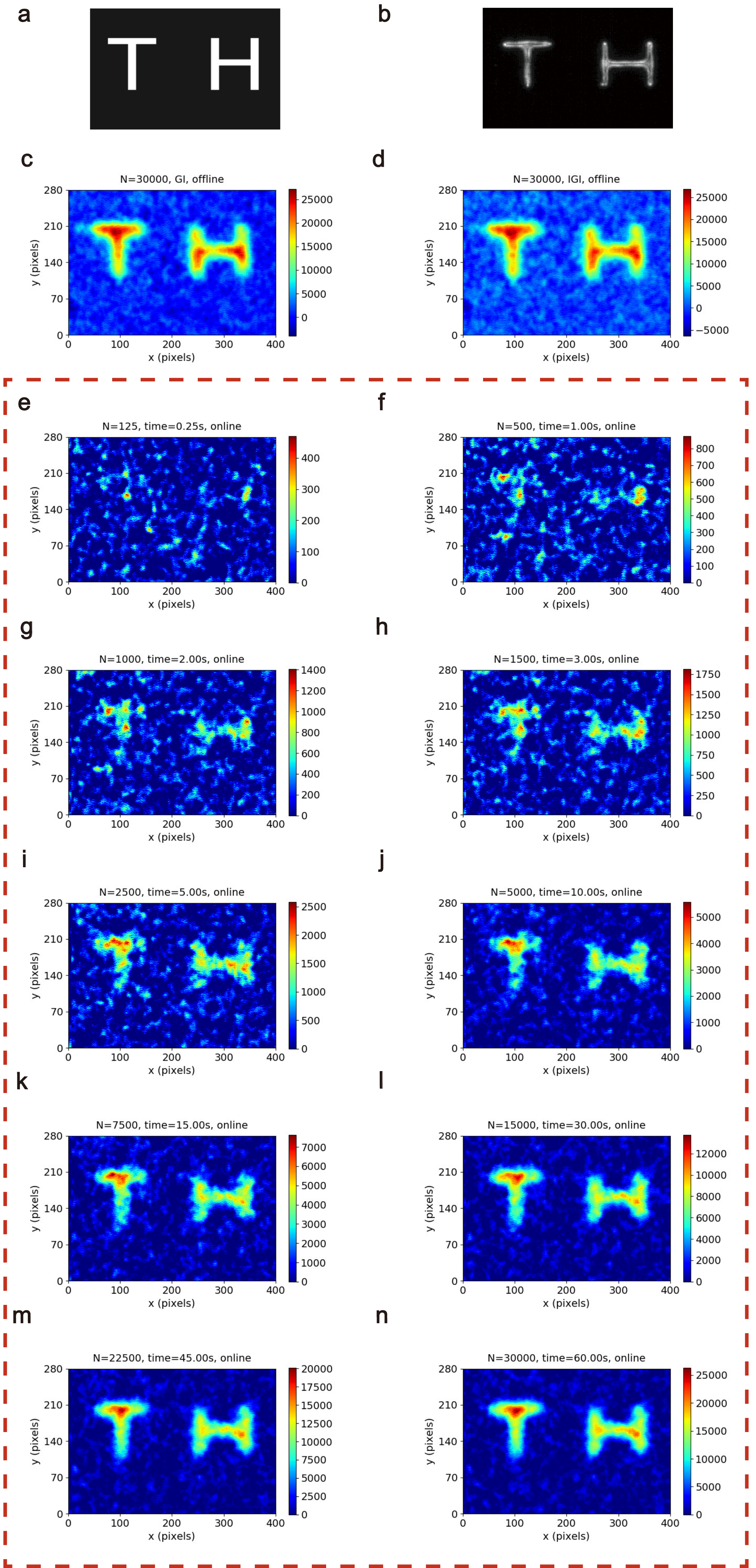}}
\caption{The images acquired by offline and online experiments. (a) The object and (b) the object imaged directly on CMOS1; the image of the object reconstituted offline by (c) the background subtraction algorithm; (d) the IGI algorithm from 30,000 measurements. (e)-(m) The intermediate results and (n) the final result of the online IGI hardware system with the number of measurements and the time in the top of each image.}
\label{figure_TH}
\end{figure}

In a similar process, we use both the offline and online methods to reconstruct the image of the object (the letters TH), which is located in the test beam extremely close to CMOS1 [Fig. \ref{figure_TH}(a)].
The image that is directly captured by CMOS1 is shown in Fig. \ref{figure_TH}(b). 
For the offline experiment, 30,000 measurements are made at a rate of 25 measurements per second, which are stored in a computer. 
The background subtraction algorithm and the IGI algorithm are used to process these data offline. 
The results, given in Figs. \ref{figure_TH}(c)-\ref{figure_TH}(d), show that the two algorithms can reconstitute a clear image of the object at a resolution of 400$\times $280.

For the online experiment, we use the on-chip IGI system to directly measure and process the data online at a rate of 500 measurements per second.
Figs. \ref{figure_TH}(e)-\ref{figure_TH}(n) show intermediate images produced by the IGI system; they show that as time increases, the ghost image gets clearer.
The image appears within 5 s after 2,500 measurements are processed by the IGI system [Fig. \ref{figure_TH}(i)]; it becomes much more finely resolved at 60 s after 30,000 measurements [Fig. \ref{figure_TH}(n)]. 
The movie shown in the monitor of the IGI system can be found in the \textcolor{urlblue}{Visualization 2}.

\subsection{Two variants of the IGI}

\begin{figure}[htb]
\centerline{\includegraphics[scale=0.33]{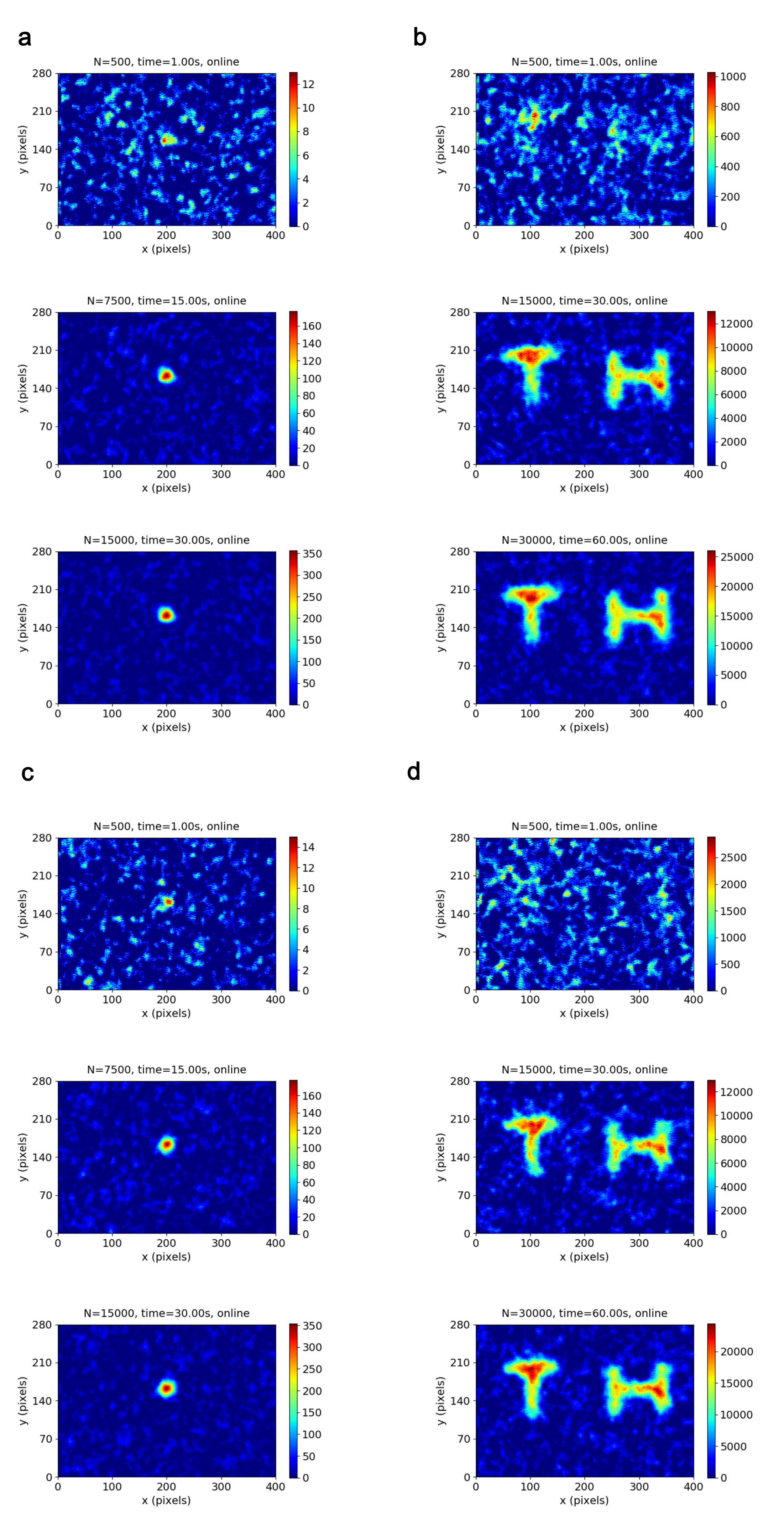}}
\caption{The experimental results of two variants. (a) The HBT effect and (b) the results of the variant  $G^{IGI_S}(x)$;  (c) the HBT effect and (d) the results of the variant $G^{IGI_I}(x)$.}
\label{figure_variants}
\end{figure}

We further propose two variants of the IGI algorithm
\begin{equation}
\begin{array}{l}
{G^{IG{I_S}}}(x) = \left\langle {({S_{n + 1}} - {S_n})[{I_{n + 1}}(x)]} \right\rangle \\
{G^{IG{I_I}}}(x) = \left\langle {({S_{n + 1}})[{I_{n + 1}}(x) - {I_n}(x)]} \right\rangle.
\end{array}
\label{IGI_S}
\end{equation}
These two variants use only the temporal differential signal of one single beam: $G^{IGI_S}(x)$ uses only the differential signals of $S$ for imaging, and $G^{IGI_I}(x)$ uses only the differential signals of $I(x)$ for imaging, rather than using both differential signals for imaging at the same time. This feature makes the algorithms easier to implement on the hardware system because fewer registers are required to store signals from only one beam. It can be demonstrated that these two variants are equivalent to the original background subtraction algorithm; the proofs are very similar to that of the original IGI algorithm. Both of these two variants have been implemented on the hardware system. The results of the HBT experiment and the GI experiment for these two variants are shown in Fig. \ref{figure_variants}. Furthermore, the two algorithms can also take the form of   $ G^{IGI_S}(x) = \left\langle {-({S_{n + 1}} - {S_n})[{I_{n}}(x)]} \right\rangle $ and $ G^{IGI_I}(x) = \left\langle {-({S_{n}})[{I_{n + 1}}(x) - {I_n}(x)]} \right\rangle $.

\subsection{Analysis of the IGI}

\begin{figure}[htb]
\centerline{\includegraphics[scale=0.35]{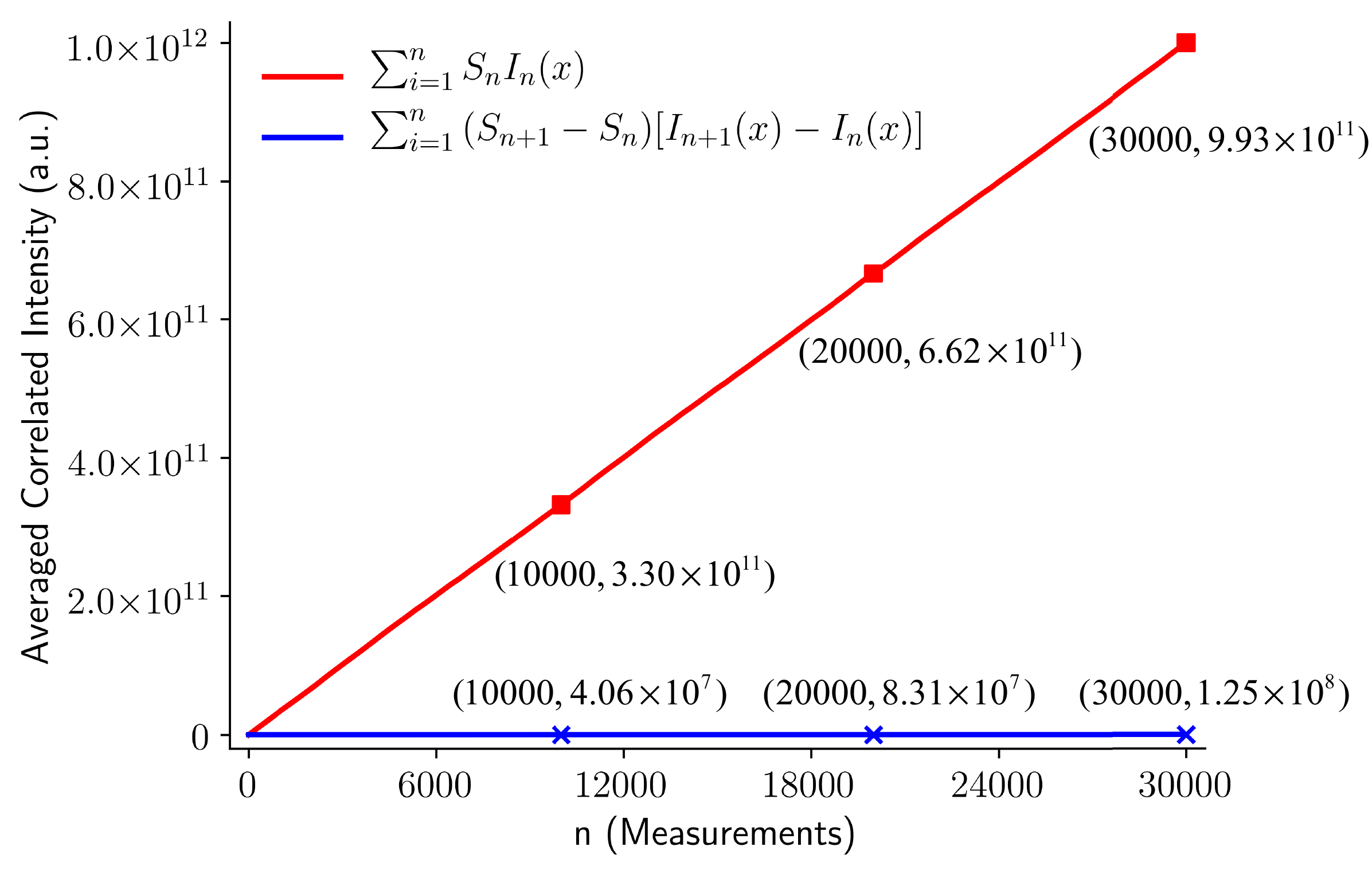}}
\caption{Analysis for the conventional GI algorithm and the IGI algorithm. Increases in the total value of $\sum\nolimits_{i = 1}^n {{S_i}{I_i}} (x)$ and $ {{\cal G}_n}(x) = \sum\nolimits_{i = 1}^n {({S_{i + 1}} - {S_i})} [{I_{i + 1}}(x) - {I_i}(x)]$  for measured $n$.}
\label{figure_incre}
\end{figure}

To determine why IGI can reduce the memory requirement and is feasible for hardware, we plotted the values of $\sum\nolimits_{i = 1}^n {{S_i}{I_i}} (x) $ and 
$ {{\cal G}_n}(x) = \sum\nolimits_{i = 1}^n {({S_{i + 1}} - {S_i})} [{I_{i + 1}}(x) - {I_i}(x)] $ as the number of measurements increased (Fig. \ref{figure_incre}). Note that the value in each case is the average of all the pixels in one image. $\sum\nolimits_{i = 1}^n {{S_i}{I_i}} (x)$  increases much more quickly than ${{\cal G}_n}(x)$. In Fig. \ref{figure_incre}, we point out that after 30,000 measurements, for a pixel on average, the GI algorithm needs to store a value of $9.93 \times {10^{11}}$ and 40 bits memory (${2^{40}} = {\rm{1.09}} \times {10^{12}}$) is needs; the IGI algorithm needs to store a value of $1.25\times {10^{8}}$, which requires 27 bits memory (${2^{27}} = {\rm{1.34}} \times {10^{8}}$). Compared with GI, IGI saves $400\times 280\times 13$ bits, which is 1.5 M-bits memory space, for a $400\times 280$ picture. The advantage of IGI algorithm is that it need not to store the meaningless average numbers of $S$ and $I(x)$, i.e. the direct current (DC) terms. Therefore, the memory space of the chip is used to store the fluctuation of thermal light. The fluctuation term of different patterns is the crucial part of the GI.

A conventional GI algorithm needs to store the values of $\sum\nolimits_{i = 1}^n {{S_i}{I_i}} (x) $, $\sum\nolimits_{i = 1}^n {{S_i}}$, and $\sum\nolimits_{i = 1}^n {{I_i}} (x) $.
However, IGI needs only to store the values of  ${{\cal G}_n}(x)$, ${{S_n}}$, and $ {{I_n}} (x)$. In Fig. \ref{figure_incre}, we just used the comparison of $\sum\nolimits_{i = 1}^n {{S_i}{I_i}} (x) $,  and ${{\cal G}_n}(x)$, to illustrate the advantages of IGI in the required memory space. In fact, for GI, the 26.9 G-bits ($30000\times 400\times 280\times 8$) memory space is needed to store 30,000 measured $I(x)$. IGI only needs 896 K-bits ($400\times 280\times 8$) bits of memory space to store one measurement of $I_n (x)$. FGPA (Xilinx XC7K325T) has only 16 M-bits of on-chip storage capacity, so only IGI can be implemented on-chip, GI cannot. Furthermore, the memory requirement of GI increases rapidly as the number of measurements increases while the memory requirement of IGI increases much more slowly, indicating that IGI needs much less memory overall.

\section{Disscusion and conclusion}

In this study, we conducted offline and online experiments to investigate both the HBT effect and lensless ghost imaging. The offline experiments validated the IGI algorithm, showing that this algorithm provides the same image quality as the background subtraction algorithm. The online experiments demonstrated the feasibility of implementing the IGI algorithm in hardware. The results show the capability of the on-chip IGI system and its two variants. The on-chip IGI system can process 500 measurements per second, and the image is reconstituted immediately after the measurement without any post-processing. Assuming that reconstructing an image with the size of 400 $\times$ 280 requires 10,000 measurements, this on-chip hardware system needs 20 s to obtain one image. We noted that there are some high-speed schemes in the field of computational GI and single-pixel imaging \cite{11hz}, for example, Xu et al. displayed a single-pixel imaging at a speed of 1,000 frames per second with the size of 32 $\times $ 32 \cite{1000hz}. However, the limiting factor of our imaging speed is the CMOS (PYTHON300, whose operating frequency is 500 measurements per second), not the speed of FPGA. Assuming that reconstructing an image requires 10,000 measurements, to reach 24 frames per second, 240,000 measurements per second need to be processed. This is not fast for the FPGA because it can operate at 100 MHz, which is 400 times faster than that is required. The measurement speed of our system can be further increased by faster CMOSs or high-speed photodiode arrays.  

The reason why the IGI can drastically reduce the memory requirement and the space complexity of GI were analyzed as follows: Firstly, the use of temporal differential signals removes the need for the space-hungry background term in the data acquisition step. Secondly, IGI requires only one frame of data from both the test and reference beams; hence, it needs less memory space to store the differential signals than conventional GI algorithms, such as a background subtraction algorithm or its normalized version. Thirdly, an empirically determined law of digital circuits states that fewer bits to be processed means fewer hardware computational resources are required to process a signal.

In summary, we have novelly developed an IGI algorithm that significantly reduced the memory requirement of conventional GI without more computational resources and degradation of image quality by using the differential signals between two consecutive temporal measurements of each beam. Although we used a lensless thermal light ghost imaging system to illustrate the capability of IGI, IGI can be directly incorporated in a CGI system. This means that IGI is applicable to GI in general.
We also conclude that the development of on-chip IGI is feasible and that all the main components, such as FPGA, CMOSs, and monitor, can be integrated onto a PCB. The on-chip implementation of IGI is significantly cheaper, smaller, and more compact than a conventional GI system, which requires computers and other digital components. These advantages pave the way for practical applications of GI. Our next step is to develop this proof-of-principle setup into a realtime imaging system that operates at more than 24 frames per second.

\section*{Appendix: Hanbury Brown and Twiss algorithm}

The conventional HBT algorithm is $
{G_{HBT}}({x_t},{x_r}) = \left\langle {[I({x_t}) - \left\langle {I({x_t})} \right\rangle ][I({x_r}) - \left\langle {I({x_r})} \right\rangle ]} \right\rangle.$
Experimentally, we can use the $N$-times measurements to calculate the HBT effect by 
\begin{equation}
{G_{HBT}}({x_t},{x_r}) = \frac{1}{N}\sum\limits_{n = 1}^N {[{I_n}({x_t}) \cdot {I_n}({x_r})] - } \frac{1}{N}\sum\limits_{n = 1}^N {{I_n}({x_t}) \cdot \frac{1}{N}} \sum\limits_{n = 1}^N {{I_n}({x_r})}.
\label{SA2}
\end{equation}

We proposed a new algorithm based on the IGI algorithm in using $(N+1)$ measurements
\begin{equation}
G_{HBT}^{IGI}({x_t},{x_r}) = \frac{1}{{2N}}\sum\limits_{n = 1}^N {[{I_{n + 1}}({x_t}) - {I_n}({x_t})]} [{I_{n + 1}}({x_r}) - {I_n}({x_r})],
\label{SA3}
\end{equation}
where ${I_{n + 1}}({x_t}) - {I_n}({x_t})$ and ${I_{n + 1}}({x_r}) - {I_n}({x_r})$ are the temporal differential signals between two consecutive measurements of the test detector and the reference detector.

We can demonstrate that the Eq. (\ref{SA3}) of the $ G_{HBT}^{IGI}({x_t},{x_r}) $ algorithm is equivalent to the Eq. (\ref{SA2}) of the conventional $ G_{HBT}({x_t},{x_r}) $  algorithm when $N$ is rather large.
It can be inferred that Eq. (\ref{SA3}) has four terms
\begin{equation}
\begin{array}{l}
G_{HBT}^{IGI}({x_t},{x_r}) = \frac{1}{{2N}}\sum\limits_{n = 1}^N {[{I_{n + 1}}({x_t}) \cdot {I_{n + 1}}({x_r})] + } \frac{1}{{2N}}\sum\limits_{n = 1}^N {[{I_n}({x_t}) \cdot {I_n}({x_r})]} \\
\;\;\;\;\;\;\;\;\;\;\;\;\;\;\;\;\;\;\;\;\;\; - \frac{1}{{2N}}\sum\limits_{n = 1}^N {[{I_{n + 1}}({x_t}) \cdot {I_n}({x_r})]}  - \frac{1}{{2N}}\sum\limits_{n = 1}^N {[{I_n}({x_t}) \cdot {I_{n + 1}}({x_r})]}. 
\end{array}
\label{SA4}
\end{equation}

When $N$ is rather large, it can be assumed that
\begin{equation}
\begin{array}{l}
\left\langle {I({x_t}) \cdot I({x_r})} \right\rangle  = \frac{1}{N}\sum\limits_{n = 1}^N {{I_n}({x_t}) \cdot {I_n}({x_r})}  \approx \frac{1}{N}\sum\limits_{n = 1}^N {{I_{n + 1}}({x_t}) \cdot {I_{n + 1}}({x_r})} \\
\left\langle {I({x_t})} \right\rangle  = \frac{1}{N}\sum\limits_{n = 1}^N {{I_n}({x_t})}  \approx \frac{1}{N}\sum\limits_{n = 1}^N {{I_{n + 1}}({x_t})} \\
\left\langle {I({x_r})} \right\rangle  = \frac{1}{N}\sum\limits_{n = 1}^N {{I_n}({x_r})}  \approx \frac{1}{N}\sum\limits_{n = 1}^N {{I_{n + 1}}({x_r})}.
\end{array}
\label{SA5}
\end{equation}
Note that two successive thermal light measurements are independent of each other. Using the statistical law that $ \left\langle {A \cdot B} \right\rangle  = \left\langle A \right\rangle  \cdot \left\langle B \right\rangle  $ when $A$ and $B$ are independent random variables together with Eq. (\ref{SA5}), we find that Eq. (\ref{SA3}) is equal to Eq. (\ref{SA2}), when N is rather large
\begin{equation}
{G_{HBT}}({x_t},{x_r}) \approx G_{HBT}^{IGI}({x_t},{x_r}).
\label{SA6}
\end{equation}
This requirement for $N$ is easy to satisfy because the number of measurement in HBT experiment is usually of the order of tens of thousands.

\section*{Funding}
National Natural Science Foundation of China (NSFC) (51727805).

\section*{Acknowledgment}
We thank prof. Kai-Li Jiang for helpful discussions. 

\section*{Disclosures}
The authors declare no conflicts of interest.

\end{document}